\documentclass[pre,preprint,showpacs]{revtex4}
\usepackage{epsfig}
\usepackage{amssymb}
\usepackage{amsfonts}
\usepackage{amsthm}
\usepackage{newlfont}
\usepackage{epsfig}
\usepackage{amscd}
\usepackage{graphics}

\newcommand{\beq}{\begin{equation}}
\newcommand{\eeq}{\end{equation}}
\newcommand{\ba}{\begin{array}}
\newcommand{\ea}{\end{array}}
\newcommand{\bea}{\begin{eqnarray}}
\newcommand{\eea}{\end{eqnarray}}
\begin{document}

\title{High probability
 state transfer in spin-1/2 chains: Analytical and numerical approaches}
\author{E.B.Fel'dman}
\email{efeldman@icp.ac.ru}
\author{E.I.Kuznetsova}
\email{kuznets@icp.ac.ru}
\author{A.I.Zenchuk}
\email{zenchuk@itp.ac.ru}

\affiliation{
Institute of Problems of Chemical Physics, Russian Academy of Sciences, Chernogolovka, Moscow reg., 142432, Russia}

\date{\today}

\begin{abstract}
This article is devoted to the
development of analytical and numerical approaches to the problem of the end-to-end quantum state transfer along the spin-1/2 chain using two methods: (a) a homogeneous spin chain with week end bonds and equal Larmor frequencies and (b) a
homogeneous spin chain with end Larmor frequencies different from inner ones. A tridiagonal matrix representation of the XY Hamiltonian  with nearest neighbor interactions   relevant to the quantum state transfer is exactly diagonalized for a combination of the above two methods. In order to take into account interactions of the remote spins we used numerical simulations of the quantum state transfer in ten-node chains. 
We compare  the state transfer times obtained using the two above  methods  for chains governed by the both XY and XXZ Hamiltonians and using both nearest neighbor and all node interactions.

\end{abstract}

\pacs{
}

\maketitle

\section{Introduction}

 The problem of  quantum state transfer  along the spin chain \cite{Bose}
acquires significance in the study of  quantum communication systems.
Several models of the end-to-end quantum state transfer along the spin-1/2 chain were suggested 
\cite{CDEL,ACDE,KS,VBR,VGIZ,GKMT,GMT,DZ}.
 All these models are based on using either  inhomogeneous chains with different coupling constants between nearest neighbors (NN) or homogeneous chains with different  Larmor frequencies on the chain nodes. Seminal works \cite{CDEL,ACDE,KS,VGIZ} use  NN interactions in the inhomogeneous symmetrical chain. All coupling constants of such chains have  fixed values, allowing one to transfer  the state during the time interval independent of the length of the chain. The method proposed in refs.\cite{VBR,VGIZ} and developed in refs.\cite{GKMT,GMT} is based on the week end bonds in the chain when interactions of all remote spins  are taken into account (the week end bonds method (WEBM)).
Transfer time increases with the length of the chain in this case. 

Since it is very difficult to construct inhomogeneous chains, an alternative method of  state transfer along the spin chain has been proposed recently \cite{DZ}. The effect of the high probability state transfer (HPST) along the homogeneous spin-1/2 chain is achieved here because of the specially tailored external inhomogeneous magnetic field.  In the simplest case one needs only two nonzero end Larmor frequencies and zero inner ones (end Larmor frequencies method (ELFM)). 

The problem of quantum state transfer  requires solving some physical questions. In particular, it is not clear whether the presence of inner nodes (so-called "body" of the chain)  improves parameters of  state transfer (transfer time and probability).  How will the transfer time be changed if we fix end nodes and remove the body? Even this very simple question requires a special study and will be considered  for two  methods of  state transfer (WEBM and ELFM) applied to the spin-1/2 chain governed by either an XY or XXZ Hamiltonian with dipole-dipole interactions.

\begin{figure*}
   \epsfig{file=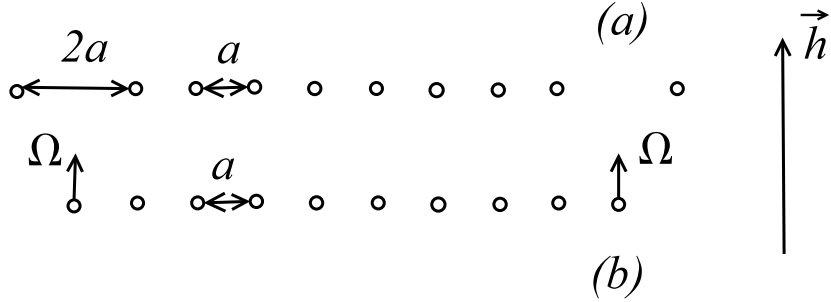
   , scale=0.4
}
  \caption{Two methods of the state transfer along the spin-1/2 chain; $a$ is the distance between nearest neighbors in the body of the chain: ($a$) WEBM, all Larmor  frequencies are zeros; ($b$) ELFM along the homogeneous chain; the end node Larmor frequencies equal to $\Omega$, while inner ones are zeros. }
  \label{Fig:methods} 
\end{figure*}

In this article, we develop an approach to the exact diagonalization of the XY-Hamiltonian with NN interactions for the combination of WEBM and ELFM, Fig.\ref{Fig:methods}. We show also the formal equivalence of the spin dynamics 
 governed by the XY and XXZ Hamiltonians in the specially tailored inhomogeneous external magnetic field in the problem  
of  quantum state transfer with a single excited spin.

We perform the numerical simulations of  quantum state transfer along the ten-node chain   with dipole-dipole interactions governed by different Hamiltonians (XY or XXZ) using different  methods of  state transfer (WEBM or ELFM) and different  types of interactions (NN or all node interactions). We concentrate on the following problems in numerical simulations:
\begin{enumerate}
\item
Study the influence of the  inner nodes on  the  transfer time over the given distance $L$. We will show that inner nodes may either descrease or increase  the transfer time in dependence on the  Hamiltonian, the method of the state transfer  and the type of interactions.
\item
Compare the end-to-end transfer times along the chains governed by 
different Hamiltonians with different  methods of the state transfer and different  types of interactions.
\end{enumerate}

Thus, the structure of this paper is following.
We give some details regarding the spin dynamics in Sec.\ref{Section:general}. 
The analytical description of the spin dynamics along the chain governed by the XY Hamiltonian with NN interactions using the combination of WEBM and ELFM is given in Sec.  \ref{Section:analitical}. The numerical simulations of the quantum state transfer  corresponding to different   Hamiltonians, methods of the state transfer and types of interactions
are represented in Sec.\ref{Section:numerics}. We briefly summarize our results in the concluding section,  Sec.\ref{Section:conclusions}.

\section{Single excited spin dynamics}
\label{Section:general}

Consider the spin dynamics along the spin-1/2 chains governed by either XY  or XXZ Hamiltonian in the 
inhomogeneous external magnetic field. These Hamiltonians read:
\begin{eqnarray}
\label{HamiltonianXY}
&&
{\cal{H}}_{XY}=\sum_{{i,j=1}\atop{j>i}}^{N}
D_{i,j}(I_{i,x}I_{j,x} + I_{i,y}I_{j,y})+ \sum_{i=1}^N \Omega^{XY}_iI_{i,z} ,
\;\;\;D_{i,j}=\frac{\gamma^2 \hbar}{2r_{i,j}^3},
\\\label{HamiltonianXXZ}
&&
{\cal{H}}_{XXZ}=\sum_{{i,j=1}\atop{j>i}}^{N}
D_{i,j}(I_{i,x}I_{j,x} + I_{i,y}I_{j,y}-2  I_{i,z}I_{j,z})+ \sum_{i=1}^N \Omega^{XXZ}_iI_{i,z} .
\end{eqnarray}
Here $D_{i,j}$ is the coupling constant between the $i$th and the $j$th nodes.
Hereafter we will study the one-dimensional spin-1/2 chains and use the dimensionless  time $\tau$, distances $\xi_{n,m}$,
 coupling constants $d_{n,m}$ and Larmor frequencies $\omega_n$,   defined as follows:
\begin{eqnarray}\label{tau}
&&
\tau= D_{1,2} t,\;\;\xi_{n,m}=\frac{r_{n,m}}{r_{1,2}},
\;\;
d_{n,m}=\frac{D_{n,m}}{D_{1,2}}=\frac{1}{\xi_{n,m}^3},\;\;d_{1,2}=1,\\\nonumber
&&
\omega^{XY}_n=
\frac{\Omega^{XY}_n}{D_{1,2}},\;\;\omega^{XXZ}_n=
\frac{\Omega^{XXZ}_n}{D_{1,2}}
.
\end{eqnarray}
Using  definitions (\ref{tau}), 
the Hamiltonians  (\ref{HamiltonianXY}) and (\ref{HamiltonianXXZ}) may be written as follows:
\begin{eqnarray}
\label{HamiltonianXYdimles}
&&
{\cal{H}}_{XY}=D_{1} \tilde {\cal{H}}_{XY},\;\;\;
\tilde {\cal{H}}_{XY}=\sum_{{i,j=1}\atop{j>i}}^{N}
d_{j,i}(I_{i,x}I_{j,x} + I_{i,y}I_{j,y})+\sum_{i=1}^N \omega^{XY}_i I_{i,z},
\\
\label{HamiltonianXXZdimles}
&&
{\cal{H}}_{XXZ}=D_{1} \tilde {\cal{H}}_{XXZ},\;\;\;
\tilde{\cal{H}}_{XXZ}=\sum_{{i,j=1}\atop{j>i}}^{N}
d_{i,j}(I_{i,x}I_{j,x} + I_{i,y}I_{j,y}-2  I_{i,z}I_{j,z})+ \sum_{i=1}^N \omega^{XXZ}_iI_{i,z}.
\end{eqnarray}

In order to characterize the effectiveness of the quantum state transfer from the first to the $N$th node, the fidelity $F(\tau)$ has been introduced  \cite{Bose}:
\begin{eqnarray}\label{fidelity}
F(\tau)=\frac{|f_{1N}(\tau)|\cos\Gamma}{3}+
\frac{|f_{1N}(\tau)|^2}{6}+\frac{1}{2},\;\;\Gamma=\arg f_{1,N}
,
\end{eqnarray}
where $f_{n,m}$ is a transfer amplitude:
\begin{eqnarray}\label{f}
f_{n,m}(\tau)&=&\langle m |e^{-i\tilde {\cal{H}} \tau}|n\rangle=\sum_{j=1}^{N}  u_{n,j} u_{m,j}e^{-i\lambda_j  \tau/2},\;\;f_{n,m}=f_{m,n}.
\end{eqnarray}
Here $u_{i,j}$, $i,j,=1,\dots,N$, are components of the normalized eigenvector
$u_j$ corresponding to the  eigenvalue $\lambda_j$ of the matrix
$D$, $H=\frac{1}{2} D$:
$Du_j=\lambda_j u_j$, where $H$ is the matrix representation of $\tilde{\cal{H}}$.

\subsection{Equivalence of XY and XXZ Hamiltonians}

It is simple to show the equivalence of  the  single excited spin dynamics governed by the XY and XXZ Hamiltonians.
In fact, two Hamiltonians (\ref{HamiltonianXYdimles}) and (\ref{HamiltonianXXZdimles}) differ 
by the diagonal parts, i.e.
\begin{eqnarray}\label{DeltaHam}
\Delta \tilde {\cal{H}} = \tilde {\cal{H}}_{XY}-\tilde {\cal{H}}_{XYZ} =
\sum_{i=1}^N( \omega^{XY}_i-\omega^{XXZ}_i)I_{i,z} + \sum_{{i,j=1}\atop{j>i}}^{N} 2 d_{i,j} I_{i,z}I_{j,z}.
\end{eqnarray}
Since we consider the single excited state transfer and both Hamiltonians commutes with $I_z$ ($z$ projection of the total spin), the spin dynamics is described by  only one block of the Hamiltonian which may be written  on the  basis of $N$ eigenvectors  $|n\rangle$, $n=1,\dots,N$, where notation $|n\rangle$ means that the $n$th node is excited, i.e. directed opposite to  the external magnetic field. Then the matrix representation  $\Delta \tilde H$ of $\Delta \tilde {\cal{H}}$  reads \cite{KF,FZ} as follows:
\begin{eqnarray}\label{DeltaHam_Matr}
&&
 \Delta \tilde{{H}} = 
 \omega^{XY}-\omega^{XXZ} - A+\Gamma E_N,\\\nonumber
&&
 \omega^{XY}={\mbox{diag}}(\omega^{XY}_1,\dots,\omega^{XY}_N),\;\;
\omega^{XXZ}={\mbox{diag}}(\omega^{XXZ}_1,\dots,\omega^{XXZ}_N),\;\;
A={\mbox{diag}}(A_1,\dots,A_N),\\\nonumber
&&
A_{n}=\sum_{{i=1}\atop{i\neq n}}^{N}d_{i,n},\;\;
\Gamma=\frac{1}{2} \sum_{{i,j=1}\atop{j> i}}^{N}d_{i,j} +\frac{1}{2}\sum_{i=1}^N (\omega^{XXZ}_i-\omega^{XY}_i).
\end{eqnarray}

Thus both Hamiltonians are equal to each other if
$\Delta \tilde H=0$. Since the number of the diagonal elements in $\Delta \tilde H$ coincides with the number of Larmor frequencies in each Hamiltonian, this condition may be satisfied  by the proper choice of the Larmor frequencies in  either the XY or XXZ Hamiltonian.
This conclusion remains correct for an arbitrary spin-1/2 system. Nevertheless, we consider XY and XXZ Hamiltonians independently as far as there is no a simple method of  generation of the inhomogeneous magnetic field. 


Remember that
   $\Gamma$ 
 may be put to zero by the proper choice of the constant homogeneous
 magnetic field \cite{Bose}. For this reason, hereafter we consider the probability of the state transfer, $P(\tau)=|f_{1N}(\tau)|^2$, instead of the fidelity as a characteristic of the state transfer effectiveness.

\section{The explicit solution for the spin dynamics in the chain governed by the $XY$-Hamiltonian  with NN interactions}
\label{Section:analitical}
In this section, we develop the ideas of Ref.\cite{KF} and give an  analytical description  of  the single excited spin  dynamics in the chain governed by the XY Hamiltonian with NN interactions  combining both WEBM and ELFM. 
In other words, one has to substitute
\begin{eqnarray}\label{delta}
d_{i,i+1}=\left\{
\begin{array}{ll}
\delta, & 1< i<N-1\cr
1, & i=1,\;N -1
\end{array}
\right., \;\;\;
\omega^{XY}_i=\left\{
\begin{array}{ll}
0, & 1< i<N\cr
\omega, & i=1,\;N 
\end{array}
\right. .
\end{eqnarray}
into the Hamiltonian (\ref{HamiltonianXYdimles}), which reads:
\begin{eqnarray}\label{HamiltonianXYdimles_NNI}
&&\tilde {\cal{H}}_{XY}=\\\nonumber
&&\sum_{i=2}^{N-2}
\frac{\delta}{2}(I^+_{i}I^-_{i+1} + I^-_{i}I^+_{i+1})+ \frac{1}{2} 
\left(I^+_{1}I^-_{2} + I^-_{1}I^+_{2} +I^+_{N-1}I^-_{N} + 
I^-_{N-1}I^+_{N} \right)+
 \omega (I_{1,z}+I_{N,z}).
\end{eqnarray}
Since we are interested in the single excited spin dynamics,  
 the Hamiltonian has $N\times N$ matrix representation (see Sec.\ref{Section:general})), which reads
(up to the scalar term, which is not important in our case and may be removed by adding a proper homogeneous   constant external magnetic field):
\begin{eqnarray}\label{D}
&&H=\frac{1}{2} D,\;\;\;
D=\left( \begin{array}{lllllr}
	2 \omega &1 & 0 & \cdots & 0& 0\\
	1& 0 & \delta& \cdots &0 & 0\\
	0 & \delta &0& \cdots &0 &0\\
	\vdots & \vdots &\vdots &\vdots &\vdots &\vdots \\
	 0& 0& 0& \cdots & 0& 1\\
	 0& 0& 0& \cdots &1& 2 \omega
\end{array}  \right).
\end{eqnarray}
First, one has to find eigenvalues and eigenvectors of $D$. For this purpose, we solve
the equation 
\begin{eqnarray}\label{Du}
D u_j = \lambda_j u_j, \;\;u_j=(u_{1j},\dots,u_{Nj})^T
\end{eqnarray}
for the components of the eigenvectors $u_j$, where $\lambda_j$ are  the solutions to the characteristic equation
\begin{eqnarray}\label{char}
\det(D-\lambda E_N)=0,
\end{eqnarray}
and $E_N$ is an $N\times N$  unit matrix.
The system (\ref{Du}) may be written as follows:
\begin{eqnarray}\label{Du1}
&&
2 \omega u_{1j} + u_{2j} =\lambda_j u_{1j},\\\label{Du2}
&&
 u_{1j} +\delta u_{3j} =\lambda_j u_{2j},\\\label{Duk}
&&
 \delta u_{(k-1)j} +\delta u_{(k+1)j} =\lambda_j u_{kj},\;\;3\le k\le N-2\\\label{DuN-1}
&&
\delta u_{(N-2)j}+u_{Nj}=\lambda_j u_{(N-1)j},\\\label{DuN}
&&
 u_{(N-1)j} + 2 \omega u_{Nj} =\lambda_j u_{Nj}.
\end{eqnarray}
Eqs.(\ref{Duk}) are known to have  the following solution
\begin{eqnarray}
&&
u_{kj}=C_{1j} e^{-i k p_j} + C_{2j} e^{i k p_j},\;\;2\le k\le N-1,\\\label{lambda}
&&
\lambda_j=2\delta \cos p_j.
\end{eqnarray}
Then eq.(\ref{char}) becomes equivalent to the compatibility condition of the system of four equations, 
eqs. (\ref{Du1},\ref{Du2},\ref{DuN-1},\ref{DuN}) for any  $j$. 
After some transformations, this compatibility condition reads
\begin{eqnarray}\label{Det2} 
&&
4i 
\left( \delta (2\omega-2\delta \cos p)\cos{\frac{N-1}{2}p}+\cos{\frac{N-3}{2}p} \right)\times\\\nonumber
&&
 \left( \delta (2\omega-2\delta \cos p)\sin{\frac{N-1}{2}p}+ \sin{\frac{N-3}{2}p} \right)
=0.
\end{eqnarray}
One can show that this equation can be treated as the $N$ degree polynomial equation for the variable $X=\cos p$, and consequently, it has $N$ roots. All roots
 can be separated into two sets of $N_1$ and $N-N_1$ roots,
which are solutions to one of two following equations:
\begin{eqnarray}\label{conditions_p} 
 &&
\delta (2\omega-2\delta \cos p_j)\cos{\frac{N-1}{2}p_j}+\cos{\frac{N-3}{2}p_j}=0, \;\;j=1,\dots, N_1,\;\;N_1=N-\left[\frac{N}{2}\right]\\ \nonumber
&&  \delta (2\omega-2\delta \cos p_j)\sin{\frac{N-1}{2}p_j}+ \sin{\frac{N-3}{2}p_j}  =0,\;\;j=N_1+1,\dots, N.
\end{eqnarray}
Here $[a]$ means the integer part of $a$.
Thus, using eqs.(\ref{conditions_p}), expressions for the normalized eigenvector components can be written as follows:
\begin{eqnarray}\label{u1}
u_{1j}&=&\left\{
\begin{array}{ll}\displaystyle
{\delta}{A_j}\cos \frac{(N-1)p_j}{2},&j=1,\dots, N_1 \cr\displaystyle
{\delta}{A_j}\sin \frac{(N-1)p_j}{2},& j=N_1+1,\dots, N
\end{array}\right.,
\\\nonumber
u_{kj}&=& \left\{
\begin{array}{ll}\displaystyle
A_j\cos \frac{(N+1-2k)p_j}{2},& k=2,...N-1,\;\; j=1,\dots, N_1\cr\displaystyle
A_j\sin \frac{(N+1-2k)p_j}{2},& k=2,...N-1,\;\;j=N_1+1,\dots, N
\end{array}\right.,
\\\nonumber
u_{Nj}&=&\left\{
\begin{array}{ll}\displaystyle
{\delta}{A_j}\cos \frac{(N-1)p_j}{2},& j=1,\dots, N_1\cr\displaystyle
-{\delta}{A_j}\sin \frac{(N-1)p_j}{2},& j=N_1+1,\dots, N
\end{array}\right.,
\end{eqnarray}
where
 \begin{eqnarray}\label{A}
A_j=\left\{
\begin{array}{ll}\displaystyle
\left( \frac{N-2}{2}+\delta^2({1+\cos (N-1)p_j})+\frac{\sin (N-2)p_j}{2\sin p_j} \right)^{-1/2},& j=1,\dots,N_1\cr
\displaystyle
\left( \frac{N-2}{2}+\delta^2({1-\cos (N-1)p_j})-\frac{\sin (N-2)p_j}{2\sin p_j} \right)^{-1/2}
,& j=N_1+1,\dots, N
\end{array}\right..
\end{eqnarray}
Then  the probability of the end-to-end excited state transfer reads:
\begin{eqnarray}\label{P}
P(\tau)&=&\left|\langle N|e^{-i{\tilde \cal{H}}_{XY}\tau}|1\rangle\right|^2=\left|\sum_{j=1}^{N}u_{Nj}u_{1j}e^{-i\tau\lambda_j/2}\right|^2=\\\nonumber
&&
{\delta^4}\left|\sum_{j=1}^{N_1}{ A_j^2}{\cos^2 (N-1)\frac{p_j}{2}}e^{-i\delta\tau\cos p_j}-\sum_{j=N_1+1}^{N}{ A_j^2}{\sin^2 (N-1)\frac{p_j}{2}}e^{-i\delta\tau\cos p_j}\right|^2,
\end{eqnarray}
which will be used in those examples of Sec.\ref{Section:numerics} where NN interactions are considered.

\subsection{Spin chain of four nodes}
The derived formulas become most simple in the case
 $N=4$. The eigenvalues (\ref{lambda}) have the  following explicit forms:
\begin{eqnarray}
\lambda_1&=&\frac{2\omega-\delta+\sqrt{(2\omega+\delta)^2+4}}{2},\;\;\;
 \lambda_2=\frac{2\omega-\delta-\sqrt{(2\omega+\delta)^2+4}}{2},\\\nonumber
\lambda_3&=&\frac{2\omega+\delta+\sqrt{(2\omega-\delta)^2+4}}{2},\;\;\;
\lambda_4=\frac{2\omega+\delta-\sqrt{(2\omega-\delta)^2+4}}{2}.
\end{eqnarray}
Formula (\ref{P}) reduces to the following one:
\begin{eqnarray}\label{probability1}
P(\tau)=\left|\frac{1}{2}e^{-i\delta\tau/4}\left\{
\cos(\tau\beta(\omega,\delta)/4)-i\frac{2\omega-\delta}{\beta(\omega,\delta)}\sin
(\tau\beta(\omega,\delta)/4) \right\} \right .\\\nonumber
-\frac{1}{2}e^{i\delta\tau/4}\left\{
\cos(\tau\alpha(\omega,\delta)/4)-i\frac{2\omega+\delta}{\alpha(\omega,\delta)}
\sin(\tau\alpha(\omega,\delta)/4) \right\}
\Big|^2,
\end{eqnarray}
where
\begin{equation}
\alpha(\omega,\delta)=\sqrt{(2\omega+\delta)^2+4},\;\; \beta(\omega,\delta)=\sqrt{(2\omega-\delta)^2+4}.
\end{equation}
Let us demonstrate that the perfect state transfer is possible at some time moment $\tau_0$ if the parameters $\omega$ and $\delta$ have been properly chosen. 
First, we remark that perfect state transfer is achieved if
\begin{eqnarray}\label{maximum1}
&&
|\cos(\tau_0\delta/4)|=1, \cos(\tau_0\alpha(\omega,\delta)/4)=\pm 1, \cos(\tau_0\beta(\omega,\delta)/4)=\mp 1,\\\nonumber
{\mbox{or}}\\\label{maximum2}
&&
|\sin(\tau_0\delta/4)|=1, \cos(\tau_0\alpha(\omega,\delta)/4)=\pm 1, \cos(\tau_0\beta(\omega,\delta)/4)=\pm 1.
\end{eqnarray}
Eqs.(\ref{maximum1}) and (\ref{maximum2})
are equivalent to the following pair of complete systems of algebraic equations (remember that $\tau_0$, $\delta$, $\alpha$, and $\beta$ must be positive, while  $\omega$ must be real):
\begin{eqnarray}\label{maximum_1}
&&
\tau_0\delta=4\pi n_1,\; \tau_0\alpha(\omega,\delta)=4\pi n_2,\; \tau_0\beta(\omega,\delta)=4\pi n_3,\\ \nonumber
&&
n_1=1,2...,\; n_2=1,2,...,\; n_2-n_3=\pm 1,\pm 3,\pm 5,..., \\ \nonumber
\text{or}\\ \label{maximum_2}
&&
\tau_0\delta=\pi (4n_1+2),\; \tau_0\alpha(\omega,\delta)=4\pi n_2,\; \tau_0\beta(\omega,\delta)=4\pi n_3, \\ \nonumber
&&
n_1=0,1,...,\; n_2=1,2...,\; n_2-n_3=0,\pm 2,\pm 4.... 
\end{eqnarray}
If $\omega=0$ (WEBM), then we obtain alternating spin chain of four nodes.  In this case, $\alpha(0,\delta)=\beta(0,\delta)=\sqrt{\delta^2+4}$ so that  system (\ref{maximum_1}) becomes inconsistent, while  system (\ref{maximum_2}) acquires the following form:
\begin{eqnarray} \label{maximum_22}
&&
\tau_0\delta=\pi (4n_1+2),\; \tau_0\alpha(0,\delta)\equiv \tau_0\sqrt{\delta^2+4}=4\pi n_2, \\ \nonumber
&&
n_1=0,1,...,\; n_2-n_1=1,2...,
\end{eqnarray}
which agrees with the results obtained for the alternating spin-1/2 chain of four nodes \cite{KF}.
The relation between $n_1$ and $n_2$ in eq.(\ref{maximum_22}) is a consequence of the inequality $\delta<\alpha(0,\delta)$.

\section{Numerical simulations of the spin dynamics.}
\label{Section:numerics}
Now we represent results of the numerical simulations of the end-to-end  single excited state transfer along the spin-1/2 chain of ten nodes. Considering the problem of  state transfer between two end nodes separated  by the dimensionless distance $L=\xi_{1,N}$,
 first, we recall the simplest model of two spins with the distance $L$ between them (Sec.\ref{Section:N2}).
Next, in Sec.\ref{Section:N10}, 
we find end-to-end transfer times over the same distance $L$ along the ten-node chain using  both XY and XXZ Hamiltonians, with both  WEBM and ELFM using both NN and all node interactions.
In particular,
we  study the problem whether the presence of inner nodes reduces the time interval required for  state transfer over the distance $L$ in comparison with the above simplest system of two interacting spins.
It will be shown that the answer is not always positive.

Note that we solve the optimization problem 
for the case of all node interactions (i.e., we find such parameters of the spin chain which provide the end-to-end HPST 
during as short as possible a time interval). Then, we consider the spin dynamics using NN interactions  and keeping the same parameters of the spin chain. This allows us to see whether NN interactions give  different  transfer times in comparison with the case of all node interactions.


\subsection{The state transfer over the distance $L$}
\label{Section:N2}
It is well known \cite{CDEL} that the  quantum state can be perfectly transferred between two  spin-1/2 nodes  separated by the distance $L=\xi_{1,N}$ during the dimensionless  time interval $\tau_2$ such that (see eq.(\ref{tau})) 
\begin{eqnarray}\label{tau_2}
 \tau_2(L)=\frac{\pi}{d_{1,N}}=L^3\pi.
\end{eqnarray}
   However the time interval $ \tau_2$  increases with an increase in $L$ so that the direct node-to-node quantum state transfer over the long distance
$L$  becomes impossible because of the quantum decoherence.  It is assumed that the set of inner nodes placed between the above two nodes may help one to  overcome the problem of decoherence. However, it seemed  that these nodes were not always helpful.

We will use the dimensionless time interval  $\tau_2(L)$ as  a characteristics of the $N$-node  chain. We say that the inner nodes in the $N$-node chain are usefull if the end-to-end transfer time along this chain is less than $\tau_2$. 

It is remarkable that the parameter $\tau_2(L)$ depends only on $L$  and does not depend on the Hamiltonians and  methods of  state transfer. 
In fact, the XY Hamiltonian coincides with the XXZ Hamiltonian up to the  scalar  term   in the matrix representation (see eq.(\ref{DeltaHam_Matr}) with  $\Omega^{XY}_1=\Omega^{XY}_N$, $\Omega^{XXZ}_1=\Omega^{XXZ}_N$ , $N=2$) as far as   the problem of the single quantum state transfer along the two-node  chain is considered. 
This   term does not effect on the probability of  state transfer.

\subsection{The ten-node  chain}
\label{Section:N10}
 
We represent results of the numerical simulations of the end-to-end state transfer along the ten-node chain using  both  WEBM and ELFM.
Considering WEBM, we will use parameter $\delta$ introduced in  eq.(\ref{delta}) and zero Larmor frequencies. Choosing the value of $\delta$, we follow the ideology of ref.\cite{GKMT}, where the distance between the end nodes and the body  of the chain is twice as long as the distance between neighbors in the body. Thus  we take $\delta=8$, see Fig.\ref{Fig:methods}a with $a=1$. However, we have  observed that all conclusions remain valid for different values of $\delta$ as well. Considering ELFM, we take $\delta=1$ and nonzero end Larmor frequencies $\omega$, while all inner Larmor frequencies are zeros, see Fig.\ref{Fig:methods}b with $a=1$ and $\Omega=\omega$.

\begin{figure*}
   \epsfig{file=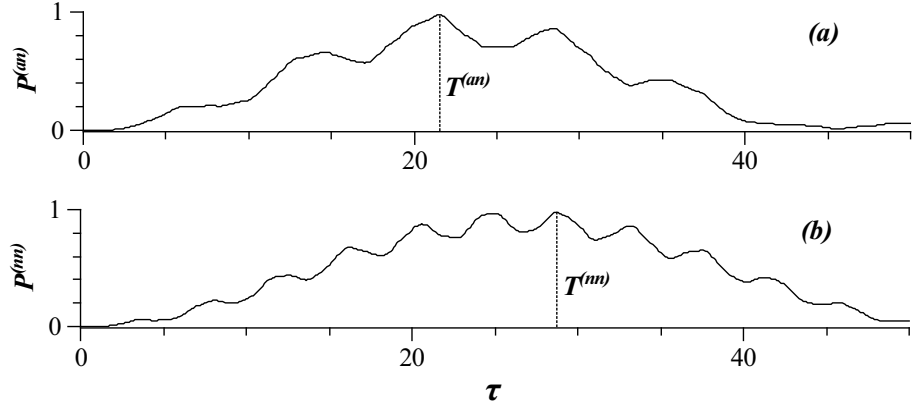
   , scale=0.7,angle=270}
  \caption{The end-to-end HPST along the chain of ten nodes governed by  the XY Hamiltonian using WEBM; $\delta=8$, $\tau_2(11/2)\approx 522.682$; (a)  all node interactions,  ${{T}}^{(an)}=21.518<\tau_2$, $P^{(an)}=0.976$;
(b) NN interactions,  ${{T}}^{(nn)}=28.698<\tau_2$, $P^{(nn)}=0.972$}
  \label{Fig:N10XYall} 
  \label{Fig:N10XYnn}                      
\end{figure*}

\begin{figure*}
   \epsfig{file=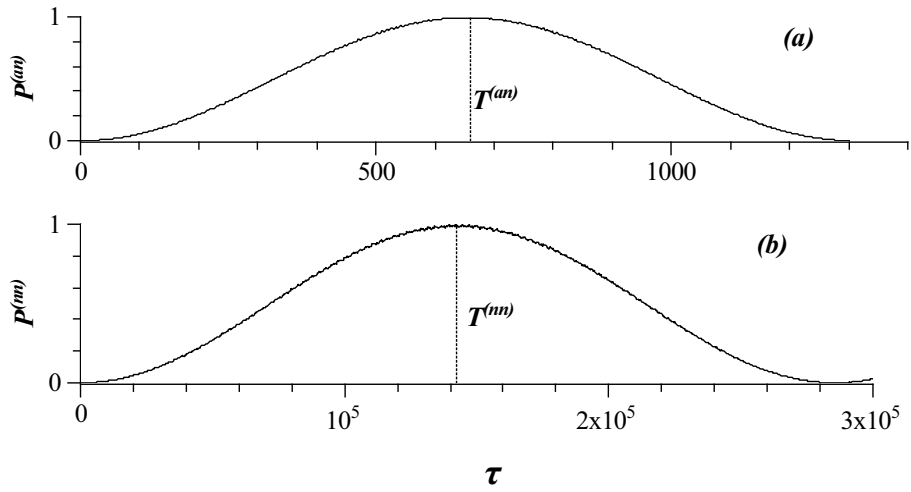
   , scale=0.7,angle=270}
  \caption{The end-to-end HPST along the chain of ten nodes governed by the XXZ Hamiltonian using WEBM; $\delta=8$, $\tau_2(11/2)\approx 522.682$; (a) all node interactions, ${{T}}^{(an)}=659.630>\tau_2$, $P^{(an)}=0.995$;
(b)  NN interactions,  ${{T}}^{(nn)}=142288.896>\tau_2$, $P^{(nn)}=1.000$}
  \label{Fig:N10XXZall} 
  \label{Fig:N10XXZnn} 
\end{figure*}

\begin{figure*}
   \epsfig{file=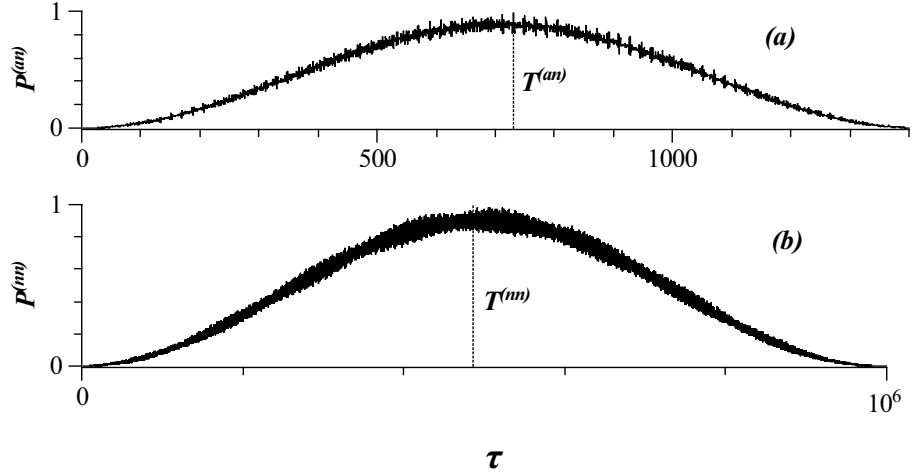
   , scale=0.7,angle=270}
  \caption{The end-to-end HPST along the chain of ten nodes  governed by the XY Hamiltonian using ELFM; $\omega=2.203$, $\tau_2(9)\approx 2290.221$; (a) 
 all node interactions,  ${{T}}^{(an)}=730.786<\tau_2$, $P^{(an)}=0.985$;
(b) NN interactions, ${{T}}^{(nn)}=485566.049>\tau_2$, $P^{(nn)}=0.994$
}
  \label{Fig:N10XYallOm} 
  \label{Fig:N10XYnnOm}
\end{figure*}

\begin{figure*}
   \epsfig{file=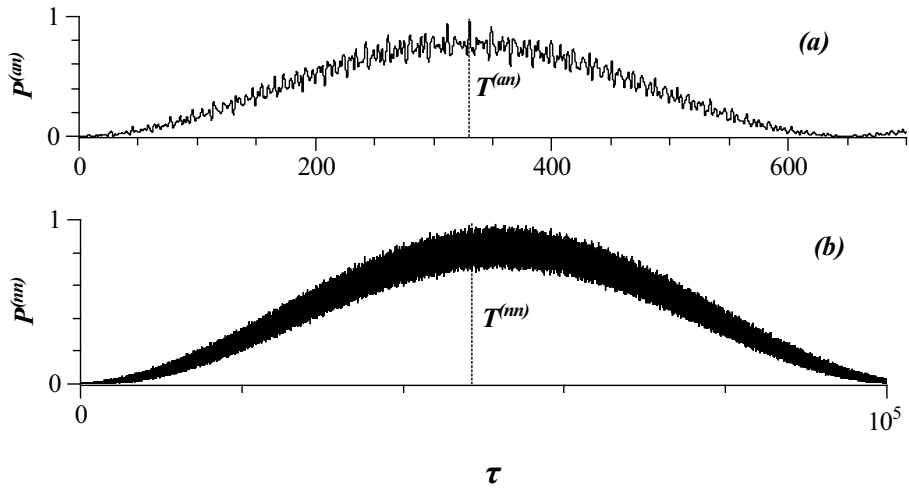
   , scale=0.7,angle=270}
  \caption{The end-to-end HPST along the chain of ten nodes with XXZ Hamiltonian and ELFM;
$\omega=2.651$, $\tau_2(9)\approx 2290.221$;
(a) all node interactions,  ${{T}}^{(an)}=330.352<\tau_2$, $P^{(an)}=0.971$;
(b) NN interactions, ${{T}}^{(nn)}=48538.313>\tau_2$, $P^{(nn)}=0.973$}
  \label{Fig:N10XXZallOm} 
  \label{Fig:N10XXZnnOm} 
\end{figure*}

Results of our numerical simulations are represented in Figs.\ref{Fig:N10XYall}-\ref{Fig:N10XXZallOm},
namely, the functions $ P(\tau)$ obtained using WEBM for the spin-1/2 chains governed by the XY and XXZ Hamiltonians are shown in Figs.\ref{Fig:N10XYall} and \ref{Fig:N10XXZall} respectively; 
similarly, functions $P(\tau)$ obtained using the ELFM for the spin chains governed by the XY and XXZ Hamiltonians  are shown in Figs.\ref{Fig:N10XYallOm} and \ref{Fig:N10XXZallOm}, respectively. Dynamics with both all node interactions
 and NN interactions are represented therein. Parameters   ${{T}}^{(an)}$, $P^{(an)}=P({{T}}^{(an)})$ and ${{T}}^{(nn)}$, $P^{(nn)}=P({{T}}^{(nn)})$ are the end-to-end   transfer times   and probabilities of these transfers in the case of    all node interactions and NN interactions  respectively.
In the following, we collect the results of analysis of all figures \ref{Fig:N10XYall}-\ref{Fig:N10XXZallOm}.

\noindent
(i) Let $T^{(WEBM)}$ and $T^{(ELFM)}$ be the transfer times corresponding to  WEBM and ELFM respectively. 
 Comparing the transfer times shown in Figs.\ref{Fig:N10XYall}-\ref{Fig:N10XXZallOm} with appropriate interval $\tau_2$ we conclude that  the inner nodes do not always reduce the transfer time in comparison with $\tau_2$ (see  Table I).
 \begin{table*}[!htb]
\begin{tabular}{|c|c|c|c|c|}
\hline
& \multicolumn{2}{c|}{all node interactions}&  \multicolumn{2}{c|}{NN interactions} \\
\cline{2-5}
    &  XY  & XXZ   &   XY & XXZ  \\\hline  
$\frac{T^{(WEBM)}}{\tau_2(11/2)}$&  0.041   &  1.262   &   0.055  &  272.228  \\
$\frac{T^{(ELFM)}}{\tau_2(9)}$&  0.319   &  0.144    &   212.017 &   21.194\\\hline
\end{tabular}
\label{Table:HPST1}
\caption{Effect of  the inner nodes on the  transfer time}
\end{table*}
We see that WEBM with an XY   Hamiltonian is the most promising method of  state transfer, namely, such chains allow one to transfer the state over the fixed distance $L$ (in this case, $\xi_{1,2}=\xi_{9,10}=1$, $\xi_{i,i+1}=1/2$, $2\le i\le 8$, so that $L=\sum_{i=1}^9\xi_{i,i+1}=11/2$)   during the shortest time interval $\frac{T^{(an)}}{\tau_2(11/2)}=0.041$ see 
 Fig.\ref{Fig:N10XYall}a and Table I.
However, one has to remember that the XXZ Hamiltonian is the most natural one describing the 
spin dynamics in the strong external magnetic field. ELFM is better suited for this Hamiltonian. It gives 
 $\frac{T^{(an)}}{\tau_2(9)}=0.144$ (in this case, $\xi_{i,i+1}=1$,  $1\le i\le 9$, so that $L=\sum_{i=1}^9\xi_{i,i+1} =9$), see Fig.\ref{Fig:N10XXZallOm}a and Table I.

\noindent
(ii) 
Comparing $T^{(an)}$ with 
$T^{(nn)}$ (which are shown in Figs.\ref{Fig:N10XYall}a-\ref{Fig:N10XXZallOm}a and \ref{Fig:N10XYall}b-\ref{Fig:N10XXZallOm}b, respectively) we 
 see that
$T^{(nn)}> T^{(an)}$ in all considered experiments with dipole-dipole interactions, which is reflected in  Table II.
\begin{table*}[!htb]
\begin{tabular}{|c|c|c|c|c|}
\hline
& {WEBM}& {ELFM} \\
\hline
 XY   & 1.334 & 664.444  \\\hline  
XXZ   &  215.710 & 146.929 \\\hline
\end{tabular}
\label{Table:HPST2}
\caption{Table of values $\frac{T^{(nn)}}{T^{(an)}}$}
\end{table*}
This observation may be explained as follows. As we have seen, the probability of the end-to-end state transfer $P(\tau)=|f_{1N}(\tau)|^2$ (where $f_{1N}$ is defined by eq. (\ref{f})) is a superposition of the oscillating functions.  Figs.\ref{Fig:N10XYall}-\ref{Fig:N10XXZallOm} demonstrate us that  the oscillation with the minimal frequency has the maximal amplitude. The transfer time is defined mainly by this oscillation. Note that  this frequency is  simply related with the minimal eigenvalue $\lambda_{min}$ of the matrix $D$ (see eq.(\ref{D})) in  the case of XY Hamiltonian with WEBM and NN interactions \cite{KZ}. In fact, $\lambda_{min}=0.118$ in this case, so that $T^{(nn)}\approx \frac{\pi}{\lambda_{min}}=26.648$, while the calculated value of $T^{(nn)}$ is $28.698$, see Fig.\ref{Fig:N10XYall}b.  Remote node interactions increase this minimal frequency,  which leads to the decrease in transfer time. 

\noindent
(iii) Let $T^{(XY)}$ and  $T^{(XXZ)}$ be  the transfer times corresponding to the chains governed by the XY and XXZ Hamiltonians respectively.  Comparing the transfer times  in Fig.\ref{Fig:N10XYall} with the appropriate times  in  Fig.\ref{Fig:N10XXZall} and the transfer times  in Fig.\ref{Fig:N10XYallOm} with the appropriate times in  Fig.\ref{Fig:N10XXZallOm}
we conclude that
\begin{eqnarray}
{\mbox{WEBM}}: && T^{(XY)} < T^{(XXZ)},\\\nonumber
{\mbox{ELFM}}: && T^{(XY)} > T^{(XXZ)},
\end{eqnarray}
see Table III.
\begin{table*}[!htb]
\begin{tabular}{|c|c|c|c|c|}
\hline
& {all node interactions}&  {NN interactions}\\\hline
WEBM & 0.033& $2.0\times 10^{-4}$\\
ELFM &2.212 & 10.004 \\\hline
\end{tabular}
\label{Table:HPST3}
\caption{Table of values $\frac{T^{(XY)}}{T^{(XXZ)}}$ }
\end{table*}

\noindent
(iv) Comparing the transfer times in Figs.\ref{Fig:N10XYall},\ref{Fig:N10XXZall} with  the appropriate times  in Figs.\ref{Fig:N10XYallOm},\ref{Fig:N10XXZallOm} we conclude that WEBM is preferable for the chains governed by the XY Hamiltonian, while ELFM is suitable for 
the XXZ Hamiltonian, see Table IV. Since we use different $L$ for WEBM ($L=11/2$) and for ELFM ($L=9$), we have to introduce parameter $b=(18/11)^3$ in order to get the correct comparison. 
 \begin{table*}[!htb]
\begin{tabular}{|c|c|c|c|c|}
\hline
&{all node interactions}&  {NN interactions} \\
\hline
     XY & 0.129 & $2.6\times 10^{-4}$                     \\\hline  

   XXZ & 8.749&  12.845     \\\hline
\end{tabular}
\label{Table:HPST4}
\caption{Table of values $\frac{T^{(WEBM)}}{T^{(ELFM)}} b$, $b=(18/11)^3$ }
\end{table*}

\section{Conclusions}
\label{Section:conclusions}

We represent an analytical and numerical approaches to the problem of single quantum state transfer along the spin-1/2 chains governed by either XY or XXZ Hamiltonian  using WEBM and ELFM. 
Let us summarize all basic results which have been obtained in this paper.

\noindent
1. We demonstrate the equivalence of the XY and XXZ Hamiltonians in the problem of the single excited state transfer along the spin chain with nonzero Larmor frequencies, i.e. one can transform the XY Hamiltonian into XXZ Hamiltonian and vice-versa taking proper Larmor frequencies.

\noindent
2. We have derived the analytical expressions for the  end-to-end state transfer probabilities along the chain governed by the XY Hamiltonian using a combination of WEBM and ELFM with NN interactions, which is most applicable to the interactions which are quickly decreasing  with an increase in the distance (like the exchange interactions).

\noindent
3. Numerical simulations of the spin dynamics along the ten-node chain with dipole-dipole interactions allow us to compare the end-to-end transfer times corresponding to the different Hamiltonians,  different methods of  state transfer, and different types of interactions, see Tables I-IV. In particular, it is shown that inner nodes are not always usefull in the process of  state transfer over the fixed distance $L$.

This work is supported by the Program of the Presidium of RAS No.7 "Development of methods of obtaining chemical compounds and creation of new materials".

\end{document}